\documentclass[prl,twocolumn,showpacs,preprintnumbers,amsmath,amssymb, superscriptaddress]{revtex4}
\usepackage{graphicx}
\usepackage{dcolumn}
\usepackage{bm}
\usepackage{fancyhdr}
\usepackage{float}
\usepackage{ifthen}
\usepackage{eurosym}
\usepackage{wrapfig}
\newcommand{\RM}[1]{\MakeUppercase{\romannumeral #1{}}}

\begin{document}
\preprint{Yb3Rh4Sn13}
\title{A neutron scattering study of the mixed state of Yb$_{3}$Rh$_{4}$Sn$_{13}$}

\ \author{D. Mazzone}
\ \affiliation{Laboratory for Neutron Scattering and Imaging, Paul Scherrer Institut, 5232 Villigen PSI, Switzerland}
\
\ \author{J. L. Gavilano}
\ \affiliation{Laboratory for Neutron Scattering and Imaging, Paul Scherrer Institut, 5232 Villigen PSI, Switzerland}

\ \author{R. Sibille}
\ \affiliation{Laboratory for Developments and Methods, Paul Scherrer Institut, 5232 Villigen PSI, Switzerland}

\ \author{M. Ramakrishnan}
\ \affiliation{Laboratory for Neutron Scattering and Imaging, Paul Scherrer Institut, 5232 Villigen PSI, Switzerland}
\ \affiliation{Swiss Light Source - Condensed Matter, Paul Scherrer Institut, 5232 Villigen PSI, Switzerland}

\ \author{M. Kenzelmann}
\ \affiliation{Laboratory for Developments and Methods, Paul Scherrer Institut, 5232 Villigen PSI, Switzerland}


\date{\today}

\begin{abstract}
Using the small angle neutron scattering (SANS) technique we investigated the vortex lattice (VL) in the mixed state of the stannide superconductor Yb$_{3}$Rh$_{4}$Sn$_{13}$. We find a single domain VL of slightly distorted hexagonal geometry for field strengths between 350 and 18500 G and temperatures between T = 0.05 and T = 6.5 K. We observe a clear in-plane rotation of the VL for different magnetic field directions relative to the crystallographic axes. We also find that the hexagonal symmetry of the VL is energetically favorable in Yb$_{3}$Rh$_{4}$Sn$_{13}$ for external fields oriented along axes of different symmetries: twofold [110], threefold [111] and fourfold [100]. The observed behavior is different from other conventional and unconventional  superconductors. The superconducting state is characterized by an isotropic gapped order parameter with an amplitude of $\Delta(0)$ = 1.57 $\pm$ 0.05 meV. At the lowest temperatures the field dependence of the magnetic form factor in our material reveals a London penetration depth of $\lambda_{L}$ = 2508 $\pm$ 17 $\AA$ and a Ginzburg coherence length of $\xi$ = 100 $\pm$ 1.3 $\AA$, i.e., it is a strongly type-II superconductor, $\kappa$ = $\lambda_{L}/\xi$ = 25. 
\end{abstract}

\pacs{74.25.Dw, 74.70.-b, 74.25.Op, 74.70.Dd}

\maketitle

\bigskip

Yb$_{3}$Rh$_{4}$Sn$_{13}$ is a member of the superconducting and/or magnetic ternary intermetallic stannides with the general formula R$_{3}$M$_{4}$Sn$_{13}$, where $R$ is an alkali metal or a rare earth and $M$ is a transition metal (Ir, Rh or Co). Several compounds of this family were first synthesized more than 30 years ago \cite{Remeika}. They crystallize in a cubic structure ($Pm$-3$n$) with a unit cell containing  40 atoms and two formula units, $Z$ = 2. Because of recent observations of interesting physical phenomena for several members of this family, the 3-4-13 stannides, they recently regained interest in the research community \cite{Klintberg, Levett}.  For instance, Ca$_{3}$Ir$_{4}$Sn$_{13}$ displays a superconducting phase with a  $T_{c}$ $\sim$ 7 K and a structural modulation with an onset at $T^*$. Under pressure $p$, $T_{c}(p)$ reveals a dome 'shaped' behavior \cite{Klintberg}. SANS results of Ca$_{3}$Rh$_{4}$Sn$_{13}$ showed a gradual change of the population of two  structural domains with a large coexistence region at 2 K \cite{Levett}.

In Yb$_{3}$Rh$_{4}$Sn$_{13}$, compared to Ca$_{3}$Rh$_{4}$Sn$_{13}$, the Ca 4$s$ electrons are replaced by the Yb 4$f$ electrons, introducing magnetism into play. In several superconducting materials the Yb atoms are in a mixed valence state near Yb$^{3+}$, as in the case of the heavy fermion systems $\alpha$- as well as $\beta$-YbAlB$_{4}$ \cite{Okawa} and YbNi$_{2}$B$_{2}$C \cite{Dhar}. Due to a hybridization of the Yb$^{3+}$ $f$-electrons with the conduction electrons, these materials display large Sommerfeld coefficients $\gamma$, indicating an enhancement of the effective mass of the itinerant charge carriers associated to the narrow $f$ bands \cite{Dhar, Nakatsuji}. Based on bond length and lattice constant analysis, it was found that the ytterbium atom in Yb$_{3}$Rh$_{4}$Sn$_{13}$ is in a mixed valence state \cite{miraglia}, with only a modest enhancement of $\gamma$ \cite{Sato}. Magnetic susceptibility studies on the superconducting state of Yb$_{3}$Rh$_{4}$Sn$_{13}$ revealed peaks in the magnetic response for specific fields below $H_{c_{2}}$ \cite{Sato}, corresponding to a broad transition into a vortex glass involving multiple steps \cite{Tomy}.

We have carried out SANS measurements in the mixed state of Yb$_{3}$Rh$_{4}$Sn$_{13}$ in the temperature range from 50 mK up to 6.5 K. The magnetic field strengths were between 350 and 18500 G and the fields were oriented along the crystallographic [100], [110] and [111] axes. We find a single domain VL of a slightly distorted hexagonal geometry. The results are very similar for all investigated temperatures and magnetic fields above 700 G. An unexpected rotation of the VL was found for magnetic field directions along the different crystallographic axes. Our data show that the superconducting state has an isotropic gapless order parameter with an amplitude of $\Delta(0)$ = 1.57 $\pm$ 0.02 meV, a $T$ = 0 K London penetration depth of $\lambda_{L}$ = 2508 $\pm$ 17 $\AA$, and a Ginzburg coherence length of $\xi$ = 100 $\pm$ 1.3 $\AA$.

The single-crystalline samples were synthesized using a Sn self-flux method. High purity elements (with atomic parts of 3 Yb, 4 Rh and a Sn flux) were heated up to 1050 $^{\circ}$C in an evacuated and sealed quartz tube. After two hours at that temperature, the liquid was cooled to 520 $^{\circ}$C with a cooling rate of 4 $^{\circ}$C/h and then quenched to room temperature. The excess tin flux was removed using diluted HCl.
\begin{figure}[tbh]
\includegraphics[width=0.8\linewidth]{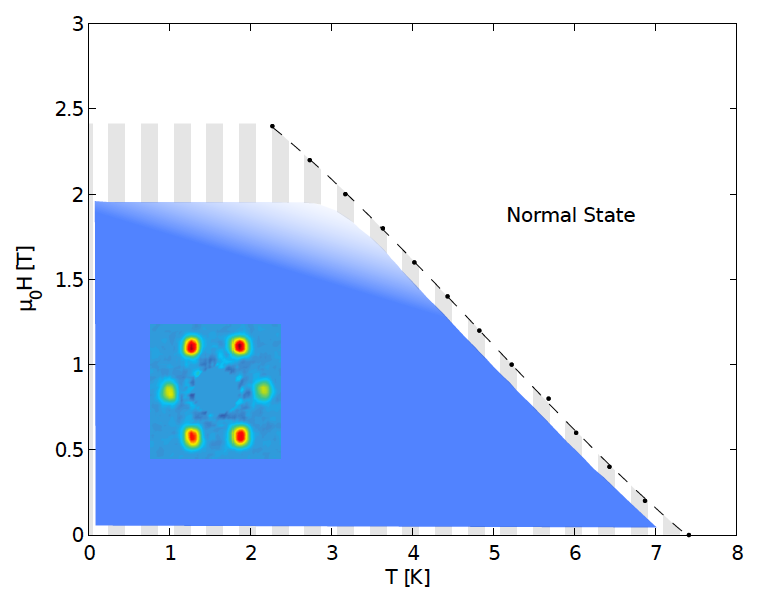}
\caption{Phase diagram of the mixed state of Yb$_{3}$Rh$_{4}$Sn$_{13}$. Electrical resistivity measurements revealed $\mu_{0}H_{c_{2}}(T)$ (black dots). The blue area indicates the region where a SANS signal was observed. In the striped area no SANS signal was observed.}
\label{phasediagramm}
\end{figure}
Electrical resistivity studies of our crystals revealed a superconducting phase transition temperature of $T_{c}$ = 7.4 K at $\mu_{0}H$ = 0 G (see Fig. \ref{phasediagramm}). In contrast to Ca$_{3}$Ir$_{4}$Sn$_{13}$, no indication for a structural modulation was found for Yb$_{3}$Rh$_{4}$Sn$_{13}$. From our heat capacity data an electronic contribution $\gamma T$, with $\gamma$ = 25.2 $\pm$ 0.2 mJ/K$^2$mol-Yb, and a phononic contribution $\beta T^3$, with $\beta$ = 1.69 $\pm$ 0.03 mJ/K$^4$mol-Yb, were extracted (data not shown). The Sommerfeld coefficient $\gamma$ is roughly four times larger than for the isostructural case Ca$_{3}$Ir$_{4}$Sn$_{13}$  \cite{Wang} with the enhancement attributed to the Yb 4$f$ bands. 

The SANS investigations were carried out on the instruments SANS-I and SANS-II at the neutron source SINQ at the Paul Scherrer Institute, Villigen, Switzerland. We used neutron wavelengths between 5.1 $\AA$ and 17.5 $\AA$ with a wavelength spread of 10\%. The incoming neutron beam was collimated over a distance of 6 m (for SANS-II), 11 m or 18 m (for SANS-I). The sample was placed into a cryomagnet with a horizontal field parallel to the incoming neutron beam. The diffracted neutron beam was detected on a two-dimensional surface detector containing of 128 x 128 pixels \cite{Kohlbrecher}. The samples were oriented by means of X-ray Laue diffraction and mounted on 0.5 mm thick Al-plates of high purity. Three different samples were used in our experiments, chosen to fit the requirements of each configuration \cite{samples}.

\begin{figure*}
\includegraphics[width=\textwidth]{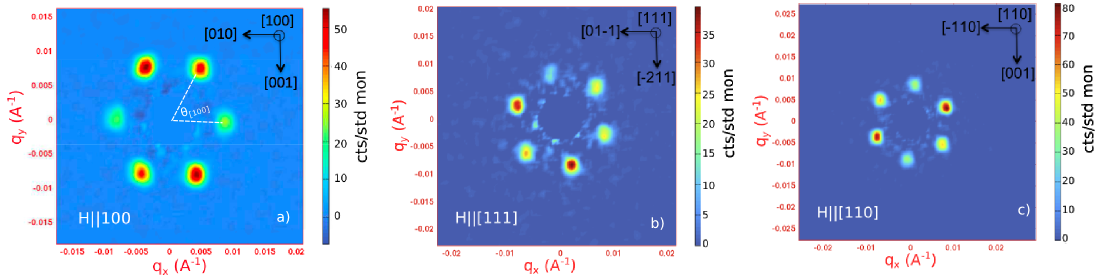}
\caption{Vortex lattice diffraction pattern at $T$ = 1.8 K and $\mu_{0}H$ = 3500 G for \textbf{a)} $\vec{H}||$[100],  \textbf{b)} $\vec{H}||$[111] and \textbf{c)} $\vec{H}||$[110].}
\label{pr}
\end{figure*}

For magnetic field strengths between 700  and 18500 G we found a single domain of a VL of nearly hexagonal structure. The obtained phase diagram is shown in Fig. \ref{phasediagramm}. The hexagonal VL symmetry usually results from an isotropic crystal structure and isotropic Fermi surface (FS). The same VL geometry was found for $H$ oriented along the three main crystallographic directions ([001], [111] and [110]), for all investigated temperatures and field strengths. It is difficult to predict the symmetry of a VL for any particular material, because it depends on details of the FS. In addition, 'pinning' of the flux lines often plays an important role. All this may favor any type of VL at a given field and temperature. One finds hexagonal, square or rhombic type of VL at low fields \cite{Muelbauer2009, Dewhurst2006}. A common approach to treat the VL uses the London model corrected for non-local effects. The latter usually trigger phase transitions between VL structures of different symmetries. In some cases one finds gradual changes of the structure with broad coexistence region of different VL symmetries \cite{Kogan1}. The latter case occurs, for instance, in the isostructural compound Ca$_{3}$Rh$_{4}$Sn$_{13}$ \cite{Levett}. We observe no temperature- or field-induced phase transitions of VL, but we can not rule out nonlocal effects.

The SANS diffraction patterns for the investigated field directions at 1.8 K and 3500 G are shown in Fig. \ref{pr}. The extracted opening angles are $\theta_{[100]}$ = 59.7 $\pm$ 0.5$^{\circ}$, $\theta_{[110]}$ = 59.9 $\pm$ 0.7$^{\circ}$ and $\theta_{[111]}$ = 59 $\pm$ 1$^{\circ}$ where the subscripts of $\theta$ denote the crystal axis along which the external field is applied. See Fig. \ref{pr} a). The values of the opening angles are consistent with a nearly ideal hexagonal lattice with only a slight distortion, if any, resulting in inequivalent $k_{x}$ and $k_{y}$ directions. The elliptical axial ratio is $\eta$ = $a/b$ = 1.07 $\pm$ 0.03, where $a$ and $b$ are the axes of the maximally distorted ellipsoid. Although the opening angles for fields along different crystallographic directions are roughly the same, varying the angle between the crystallographic axes and the external field results in an in-plane rotation of the VL with respect to the crystal orientation. As shown in Fig. \ref{pr} a) and c), the magnetic Bragg spots for $\vec{H}||$[100] are turned by 22 $\pm$ 1$^\circ$ compared to the spots for $\vec{H}||$[110]. In both cases the [001] axis is vertical. To our knowledge, this rotation is rarely observed in other superconductors. A possible scenario to explain this observation may involve the explicit consideration of a rotational modulus $Cr$ as in Ref. \cite{Miranovic}.

\begin{figure*}[tbh]
\includegraphics[width=0.8\textwidth]{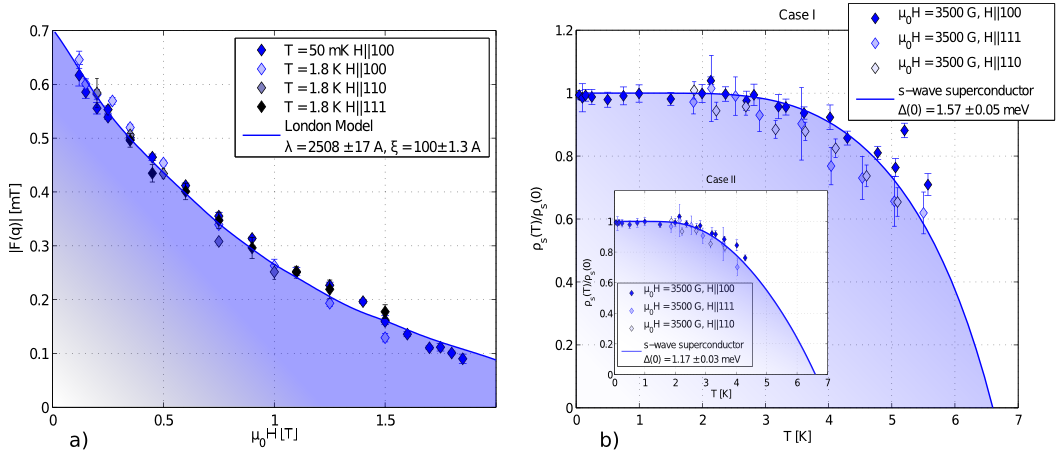}
\caption{\textbf{a)} Field dependence of the magnetic form factor for the three crystal directions at $T$ = 50 mK and $T$ = 1.8 K. \textbf{b)} Temperature dependence of the superfluid density. The data reveal a $s$-wave superconducting order parameter and $\Delta(0)$ = 1.57$\pm$ 0.06 meV. Inset: Temperature dependence of the superfluid density neglecting the temperature dependence of $\xi$ with $\Delta(0)$ = 1.17$\pm$ 0.03 meV. Solid lines represent best fit to the data (see text).}
\label{gapanalysis}
\end{figure*}

From the VL diffraction patterns the characteristic properties of the superconducting state, such as the London penetration depth $\lambda_{L}$, the Ginzburg coherence length $\xi$ and the absolute value of the superconducting order parameter are obtained as follows. The VL of a type-II superconductor forms a two-dimensional crystal in $q$-space. The local field in the mixed state can be written as a sum over spatial Fourier components called the magnetic form factor F($\vec{q}_{h,k}$). The integrated intensity $I_{int}$($\vec{q}_{h,k}$) of a magnetic Bragg spot is then
\begin{equation}
	I_{int}(\vec{q}_{h,k})=2\pi V \phi_{n}\Big(\frac{\mu_{n}}{4}\Big)^2\frac{\lambda_{n}^2}{\Phi_{0}^2q_{h,k}cos(\zeta)}|F(\vec{q}_{h,k})|^2,
	\label{intin}
\end{equation}
 where V is the volume of the sample, $\phi_{n}$ is the incident neutron flux per unit area, $\lambda_{n}$ the wavelength of the incident neutron beam, $\mu_{n}$ = 1.91 the magnetic moment of the neutron in nuclear magnetons and $\Phi_{0}$ the flux quantum \cite{Christen}. The integrated intensity is recorded from a Gaussian fit over the rocking curve of each Bragg spot as a function of the rocking angle. This rotation is taken into account in Eq. \ref{intin} by the Lorentz factor cos($\zeta$). Here, $\zeta$ represents the angle between the reciprocal lattice vector and the direction normal to both (i) the rotation axis of the cryomagnet and (ii) the incoming neutron direction. The field dependence of the magnetic form factor of Yb$_{3}$Rh$_{4}$Sn$_{13}$ is shown in Fig. \ref{gapanalysis}. There is no significant difference in the field dependence of the form factor for the different field directions relative to the crystal. This result hints for isotropic superconducting properties. Furthermore, no significant difference of the field dependent F($\vec{q}_{h,k}$) is observed between 50 mK and 1.8 K.
 
From the magnetic field dependence of F($\vec{q}_{h,k}$), $\lambda_{L}$ and $\xi$ are obtained using a London model that takes into account the finite size of the vortex cores. The magnetic form factor is then given by \cite{Eskildsen}:
 
   \begin{equation}
	F(\vec{q}_{h,k})=\frac{Be^{-0.44q^2\xi^2}}{1+q_{h,k}^2\lambda_{L}^2},
	\label{londonvortex}
\end{equation} 
where $B$ is the applied magnetic field and $\vec{q}_{h,k}$ the field-dependent reciprocal lattice vector directly measured in our SANS experiments. The magnetic field dependence of the magnetic form factor of Yb$_{3}$Rh$_{4}$Sn$_{13}$ was fitted using Eq. \ref{londonvortex} with $\lambda_{L}$ and $\xi$ as free parameters. The best fit to the data reveals a London penetration depth of $\lambda_{L}$ = 2508 $\pm$ 17 $\AA$ and a Ginzburg coherence length of $\xi$ = 100 $\pm$ 1.3 $\AA$. From the ratio of these two values, the Ginzburg-Landau (GL) parameter $\kappa$ = $\lambda_{L}/\xi$ = 25 $\pm$ 0.3 is obtained, indicating a strong type-II superconductor. The same results are obtained for fields oriented along [100], [111] and [110] axis. Thus, $\lambda_{L}$ and $\xi$ are the same along all the main directions of the crystal.
 
For the nonmagnetic superconductor Ca$_{3}$Rh$_{4}$Sn$_{13}$, a GL parameter of 6.3 was  inferred from SANS data ($\lambda_{L}(0)$ $\approx$ 841 $\AA$) \cite{Levett}. Replacing Ca by the rare earth Yb the GL parameter is increased by a factor of four. The larger London penetration depth ($\lambda_{L}\propto\sqrt{m^*/n_{s}}$) of Yb$_{3}$Rh$_{13}$Sn$_{13}$ results from an enhancement of the effective mass or a suppression of the superfluid density. This attributed to differences of the FS due to the 4$f$ Yb bands. 

We recorded several Bragg reflections at various temperatures above $T$ = 50 mK and found no indications for temperature-dependent changes in the position and the longitudinal FWHM ($\Gamma_{L}$). $\Gamma_{L}$ is normally measured by recording the SANS signal in the rocking curve as a function of the rocking angle for each Bragg peak but this could be done more efficiently in our case \cite{tempde}. To extract the temperature dependence of $\lambda_{L}$ from the collected Form factors we have tried two very different approximations for $\xi(T)$: (\RM{1}) We assume that the coherence length follows a weak-imit BCS behavior, $i.e.$, $\xi(T)$ = $\xi(0)[$tanh$(1.78\sqrt{T_{c}/T-1})]^{-1}$ with $\xi(0)\approx\xi$(50mK) in Eq. \ref{londonvortex}. In a second type of approximation (\RM{2}) we simply neglect the temperature dependence of the coherence length. In case \RM{1} fits were implemented for $T<$ 0.85$T_{c}$.  Approximation \RM{2}, of course, has stronger limitations. The fits were extracted to $T\leq$ 0.6$T_{c}$. In either case,
 \begin{equation}
\rho_{s}(T)/\rho_{s}(0)=\lambda_{L}(0)^2/ \lambda_{L}(T)^2, 
\label{SFD}
\end{equation}

with $\lambda_{L}(0)\approx\lambda_{L}$(50mK) and shown in Fig. \ref{gapanalysis} b) for 3500 G ($\vec{H}||$[100], $\vec{H}||$[110] and $\vec{H}||$[111]). Results above $T$ = 0.85$T_{c}$ ($T$ = 0.6$T_{c}$, respectively) are not shown and are not taken into account in the fits. 

The temperature dependent superfluid density for a system with a three-dimensional Fermi surface and an isotropic $s$-wave pairing may be approximated by \cite{Prozorov}
 \begin{equation}
\frac{\rho_{s}(T)}{\rho_{s}(0)}=1-\frac{1}{2T}\int_{0}^{\infty}\textrm{cosh}^{-2}\Bigg(\frac{\sqrt{\varepsilon^2+\Delta^2(T)}}{2k_{B}T}\Bigg)d\varepsilon \textrm{.}
\label{gapswave}
\end{equation}
Here $T$ denotes the temperature, $k_{B}$ is the Boltzman constant, $\sqrt{\varepsilon^2+\Delta^2(T)}$ the excitation energy of the superconducting state. The temperature dependence of the superconducting order parameter is approximated by \cite{White2}
 \begin{equation}
\Delta(T)\cong\Delta(0)\textrm{tanh}(1.78\sqrt{T_{c}/T-1})\textrm{,}
\label{DeltaT}
\end{equation}
where $T_{c}$ was fixed from results of thermodynamic measurements preformed on the same samples (see Fig. \ref{phasediagramm}). The best fit to the data using Eq. \ref{gapswave} and approximation \RM{1} - the most reliable - yields $\Delta(0)$ = 1.57 $\pm$ 0.05 meV and the coupling strength $\alpha$ = $\Delta(0)/k_{B}T_{c}$ = 2.8 $>$ 1.76. The same conclusion results from the analysis obtained at different fields, 3500, 7500 and 12500 G. We conclude that Yb$_{3}$Rh$_{4}$Sn$_{13}$ is a type-II superconductor with a strong electron-phonon coupling. For completeness we mention that using the approximation \RM{2} to extract $\Delta(T)$ from Eq. \ref{londonvortex} we find $\alpha$ = $\Delta(0)/k_{B}T_{c}$ = 2.1 (see insert of Fig. \ref{gapanalysis} b). The relative difference in the superconducting order parameter is of the order of 25\%. This seems reasonable for the used approximations. While approximation \RM{1}, most likely, overcorrects the temperature dependence of the coherence length it is, nevertheless, the most reliable. Approximation \RM{2} reveals a lower boundary of the superconducting order parameter.
\begin{figure}[tbh]
\includegraphics[width=0.8\linewidth]{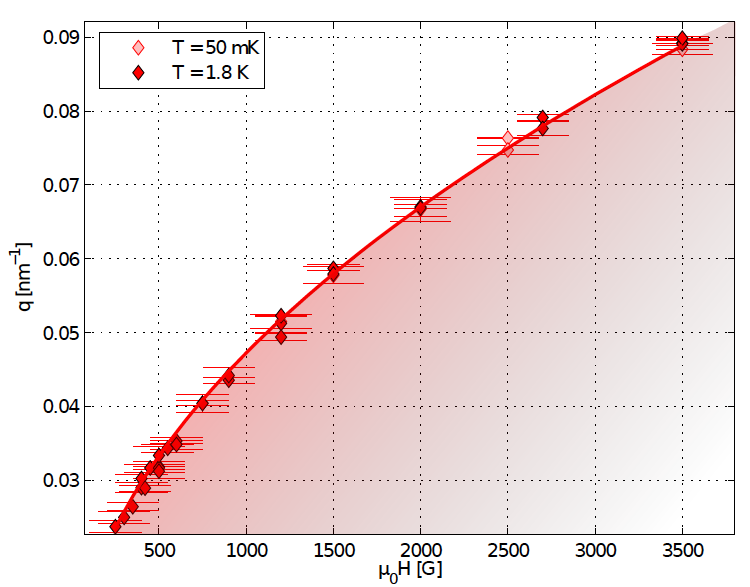}
\caption{$q$ dependence of the VL following the typical square-root behavior.}
\label{fwhm}
\end{figure}

In our experiments the demagnetization field $\Delta B$, estimated from reported magnetization results \cite{Sato}, was less than 3\% and thus, neglected. Nevertheless, for decreasing magnetic field strengths below approximately 7500 G a gradual increase of the FWHM of the rocking curve $\Gamma_{L}$ is observed (see Ref. \cite{DanielMelting}). The increase of $\Gamma_{L}$ at smaller magnetic field strengths is independent on the crystal orientation. Furthermore, no evidence for a thermal-fluctuation-induced increase of $\Gamma_{L}$  was found between 50 mK and 5 K. Despite the increasing broadening of the magnetic Bragg reflections their position follows the typical square-root behavior for a hexagonal VL down to magnetic field strengths of 350 G (see Fig.  \ref{fwhm}). Therefore, we concluded that there are no temperature- or field-induced VL transitions involving changes of the geometry. The rapid increase of the width $\Gamma_{L}$ at low, but also at high fields (data not shown) hits for a melting of the VL as discussed in Ref. \cite{DanielMelting}.

In summary, we have investigated the mixed state of Yb$_{3}$Rh$_{4}$Sn$_{13}$ by means of SANS. We found a single domain of a nearly ideal hexagonal VL structure for fields between 700 and 18500 G and temperatures between 50 mK and 6.5 K. Changes of the field orientation, relative to the crystal, yield no sign for a change of the symmetry of the VL. We find only an unexpected in-plane rotation of the vortex lattice for different crystal directions relative to the external field. It is unusual in a cubic system that a single VL symmetry prevails for all the field directions and field strengths. Temperature fluctuations of the VL seem to play no significant role on the VL static properties. Yb$_{3}$Rh$_{4}$Sn$_{13}$ can be characterized as a type-II superconductor in the strong coupling limit with $\lambda_{L}$ = 2508 $\pm$ 17 $\AA$ and $\xi$ = 100 $\pm$1.3 $\AA$. A coupling parameter of $\Delta(0)/k_{B}$ = 2.8 was found in the clean limit of superconductivity.

We thank the Swiss National Foundation for the support of D. M. (project number 200021\_147071). R. S. is supported by the European Community's Seventh Framework (FP7/2007-2013), grant number 290605 and COFUND: PSI-FELLOWSHIP. M.R. has been supported by the Erasmus Mundus program MaMaSELF.

\bibliographystyle{apsrev}

\bibliography{yb3rh4sn13}

\begin{thebibliography}{20}
\bibitem[{\citenamefont{samples}(200)\citenamefont{samples}}]{samples}
\bibinfo{author}{\bibfnamefont{The samples aligned with the external field and the incoming neutron beam along [001] and [111] direction were  single crystals of 294 mg and 179 mg. The sample, aligned along the [110] consisted of 3 co-aligned samples of 50 mg each.}}
\bibitem[{\citenamefont{tempde}(200)\citenamefont{tempde}}]{tempde}
\bibinfo{author}{\bibfnamefont{Our data evidence that the structure of the VL is at most only weakly temperature dependent. Thus, in our measurements of the Bragg peaks, the integrated intensity is calculated from the detected peak intensity and the previously determined $\Gamma_{L}$ for some chosen temperatures, i.e., $I_{int}$ $\approx$ $I_{peak}\Gamma_{L}$.}}
\end{thebibliography}

\end{document}